\documentclass[12pt,preprint]{aastex}
\usepackage{emulateapj5}
\newcommand{\mincir}{\raise
-3.truept\hbox{\rlap{\hbox{$\sim$}}\raise4.truept\hbox{$<$}\ }}
\newcommand{\magcir}{\raise
-3.truept\hbox{\rlap{\hbox{$\sim$}}\raise4.truept\hbox{$>$}\ }}
\newcommand{\minmag}{\raise
-3.truept\hbox{\rlap{\hbox{$<$}}\raise5.truept\hbox{$<$}\ }}
\newcommand{\be}{\begin{equation}}
\newcommand{\ee}{\end{equation}}
\newcommand{\ba}{\begin{eqnarray}}
\newcommand{\ea}{\end{eqnarray}}
\newcommand{\brr}{\begin{array}}
\newcommand{\err}{\end{array}}
\newcommand{\bc}{\begin{center}}
\newcommand{\ec}{\end{center}}

\makeatother

\shorttitle{Luminosity Dependent X-ray AGN Clustering}
\shortauthors{M. Plionis et al.}

\begin{document}

\title{Luminosity Dependent X-ray AGN Clustering ?}

\author{M. Plionis,\altaffilmark{1,2}, M. Rovilos\altaffilmark{1,3},  
S. Basilakos\altaffilmark{4}, I. Georgantopoulos\altaffilmark{1}, 
F. Bauer\altaffilmark{5} }
\altaffiltext{1}{Institute of Astronomy \& Astrophysics, 
National Observatory of Athens, I.Metaxa \& B.Pavlou, 
P.Penteli 152 36, Athens, Greece}
\altaffiltext{2}{Instituto Nacional de Astrofisica, 
Optica y Electronica (INAOE)
Apartado Postal 51 y 216, 72000, Puebla, Pue., Mexico}
\altaffiltext{4}{currently at: Max-Planck-Institut für extraterrestrische 
Physik, Giessenbachstaße, Postfach 1312, 85748 Garching, Germany}
\altaffiltext{4}{Academy of Athens, Research Centre for Astronomy \& Applied
      Mathematics, Soranou Efessiou 4, 11-527 Athens, Greece}
\altaffiltext{5}{ Columbia Astrophysics Laboratory, Columbia University, 
Pupin Physics Laboratories, 550 West 120th St., New York, USA}

\begin{abstract}
We have analysed the angular clustering of X-ray selected active 
galactic nuclei (AGN) in different
flux-limited sub-samples of the {\em Chandra} Deep Field North (CDF-N) 
and South (CDF-S) surveys. We find a strong dependence of the
clustering strength on the sub-sample flux-limit, a fact which explains most of
the disparate clustering results of different XMM and {\em Chandra} 
surveys. 
Using Limber's equation, we find that the inverted CDF-N and 
CDF-S spatial clustering lengths are consistent with direct spatial clustering
measures found in the literature,
while at higher flux-limits the clustering length increases
considerably; for example, at
$f_{x,{\rm limit}}\sim 10^{-15}$ erg s$^{-1}$ cm$^{-2}$ 
we obtain $r_0\simeq 17 \pm 5$ and $18\pm 3 \; h^{-1}$ Mpc, for the 
CDF-N and CDF-S, respectively.
We show that the observed flux-limit clustering trend hints towards 
an X-ray luminosity dependent clustering of X-ray selected, $z\sim 1$, AGNs. 

\end{abstract}


\section{Introduction}
X-ray selected AGNs provide a relatively unbiased census of the 
AGN phenomenon, since obscured AGNs, largely missed in optical surveys, 
are included in such surveys.
Furthermore, they can be detected out to high redshifts and thus trace
the distant density fluctuations providing important
constraints on supermassive black hole formation, 
the relation between AGN activity and Dark Matter (DM) halo hosts,
the cosmic evolution of the AGN phenomenon (eg. Mo \& White 1996, 
Sheth et al. 2001), 
and on cosmological parameters and the dark-energy 
equation of state (eg. Basilakos \& Plionis 2005; 2006, Plionis \& Basilakos 
2007).

Until quite recently our knowledge of X-ray AGN clustering came
exclusively from analyses of ROSAT data ($\le 3$keV) 
(eg. Boyle \& Mo 1993; Vikhlinin \& Forman 1995; Carrera et al. 1998; 
Akylas, Georgantopoulos, Plionis, 2000; Mullis et al. 2004). 
These analyses
provided conflicting results on the nature of high-$z$ AGN 
clustering. Vikhlinin \& Forman (1995), using the angular correlation approach
and inverting to infer the spatial correlation length, found a strong amplitude
of $\bar{z}\sim 1$ sources ($r_{0}\simeq 9 \; h^{-1}$ Mpc), which translates
into $r_{0}\simeq 12 \; h^{-1}$ Mpc for a 
$\Lambda$CDM cosmology and a luminosity 
driven density evolution (LDDE) luminosity function (eg. Hasinger et al. 2005).
Carrera et al. (1998), however, 
%investigated the spatial clustering directly 
using spectroscopic data, could not confirm such a 
large correlation amplitude.% for the high-$z$ X-ray AGNs.

With the advent of the XMM and {\em Chandra} X-ray observatories, 
many groups have attempted to settle this
issue. Recent determinations of the high-$z$ X-ray selected AGN 
clustering, in the soft and hard bands, have provided 
again a multitude of conflicting results, intensifying the debate
(eg. Yang et al. 2003; Manners et al. 2003;
Basilakos et al. 2004; Gilli et al. 2005; Basilakos et al 2005; 
Yang et al. 2006; Puccetti et al. 2006; Miyaji et al. 2007; Gandhi et al.
2006; Carrera et al. 2007).

In this letter we investigate these clustering 
differences by re-analysing the CDF-N and CDF-S surveys,
using the Bauer et al. (2004) classification to select 
only AGNs in the 0.5-2 and 2-8 keV bands. 
To use all the available sources, and not only those having
spectroscopic redshifts, we work in 
angular space and then invert the angular correlation
function using Limber's equation.
Hereafter, we will be using $h\equiv H_{\circ}/100$ km $s^{-1}$ Mpc$^{-1}$.

\section{ The X-ray source catalogues} 
The 2Ms CDF-N and 1Ms CDF-S {\it Chandra} data  represent the deepest 
observations currently available at X-ray wavelengths (Alexander et al. 2003, 
Giaconni et al. 2001). 
The CDF-N and CDF-S cover an area of 448 and 391 arcmin$^{2}$, 
respectively. We use the source catalogues of Alexander et al. (2003) 
for both CDF-N and CDF-S. 
The flux limits that we use for the CDF-N 
 are $3\times 10^{-17}$ and $2\times 10^{-16}$  
 $\rm erg~cm^{-2}~s^{-1}$ in the soft and hard band, 
 while for the CDF-S the respective values are 
 $6\times 10^{-17}$ and $5\times 10^{-16}$ erg cm$^{-2}$ s$^{-1}$. 
Note that sensitivity maps were produced following the prescription of 
Lehmer et al. (2005) and in order to produce random catalogues in a 
consistent manner to the source selection, we discard sources which 
lie below our newly determined 
sensitivity map threshold, at their given position. 
Our final CDF-N catalogues contain 383 and 263 sources 
in the soft (0.5-2 keV) and hard band (2-8 keV), 
respectively, out of which 304 and 255 are AGNs, according to
the ``pessimistic'' Bauer et al. (2004) classification. 
The corresponding CDF-S catalogues
contain 257 and 168 sources in the same bands, out of which 227 and
165 are AGNs. 
A number of sources (roughly half) have spectroscopic redshift determinations 
(mostly taken from Barger et al. 2003; 
Szokoly et al. 2004; Vanzella et al. 2005; Vanzella et al. 2006; 
Le F\'evre et al. 2004; Mignoli et al. 2005).

\section{Correlation function analysis}
\subsection{The angular correlation}
The clustering properties of the X-ray AGNs are
estimated using the two-point angular correlation function, $w(\theta)$,
estimated using $w(\theta)=f(N_{DD}/N_{DR})-1$,
where $N_{DD}$ and $N_{DR}$ is the number of data-data and data-random
pairs, respectively, within separations $\theta$ and  $\theta+d\theta$. 
The normalization factor is given by
$f = 2 N_R /(N_D-1)$, where $N_D$ and $N_R$ are the total number of
data and random points respectively. The Poisson uncertainty in  
$w(\theta)$ is estimated as $\sigma_{w}=\sqrt{(1+w(\theta))/N_{DR}}$
(Peebles 1973).
 
The random catalogues are produced to account for the different 
positional sensitivity and edge effects of the surveys. To this end
we generated 1000 Monte Carlo random realizations of the 
source distribution, within the CDF-N and CDF-S survey areas, 
by taking into account the local variations in sensitivity.
We also reproduce the desired $\log N - \log S$ distribution, 
either the Kim et al. (2007) or the one recovered 
directly from the CDF data (however our results remain mostly unchanged 
using either of the two).
Random positioned sources with fluxes lower 
than that corresponding to the particular position of the sensitivity
map are removed from our final random catalogue. 

We apply the correlation analysis evaluating $w(\theta)$ in the range
$[5^{''}, 900^{''}]$ in 10 logarithmic intervals with $\delta \log
\theta= 0.226$. 
We find statistically significant signals for all bands and for both
CDF-N and CDF-S.  As an example we provide in Table 1 the integrated signal 
to noise ratios, given for two different flux-limits and two different 
angular ranges.
The significance appears to be 
low only for the CDF-S hard-band, but for the lowest flux-limit. 

The angular correlation function for two different flux-limits 
are shown in Figure 1, with the lines corresponding to the best-fit 
power law model: $w(\theta)=(\theta_{0}/\theta)^{\gamma-1}$,
using $\gamma=1.8$ and the standard $\chi^{2}$ 
minimization procedure. Note that for the CDF-N, we get 
at some $\theta$'s very low or negative  
$w(\theta)$ values. These, however, are taken into account
in deriving the integrated signal, presented in Table 1.

Applying our analysis for different flux-limited sub-samples, we find 
that the clustering strength increases with increasing flux-limit, in
agreement with the CDF-S results of Giacconi et al (2001). 
In Figure 2 we plot the angular clustering scale, $\theta_0$, derived from
the power-law fit of $w(\theta)$, as a function of different sample
flux-limits. The trend is true for both energy bands and  
for both CDF-N and CDF-S, although for the latter is apparently
stronger.

We also find that at their lowest respective flux-limits the 
clustering of CDF-S
sources is stronger than that of CDF-N (more so for the soft-band), in 
agreement with the spatial clustering analysis of Gilli et al. (2005). This
difference has been attributed to cosmic variance, in the sense that 
there are a few
large superclusters present in the CDF-S (Gilli et al. 2003).
However, selecting CDF-N and CDF-S sources at the same flux-limit reduces 
this difference, which remains strong only for the highest flux-limited 
sub-samples (see Fig.2).

It is worth mentioning that our results could in principle
suffer from the so-called {\em amplification bias}, which can enhance
artificially the clustering signal due to the detector's PSF smoothing 
of source pairs with intrinsically small angular separations
(see Vikhlinin \& Forman 1995; Basilakos et al. 2005). 
However, we doubt whether this bias can significantly affect our results
because at the median redshift of the sources ($z\sim 1$) the {\em Chandra}
PSF angular size of $\sim 1^{''}$ corresponds to a rest-frame
spatial scale of only $\sim 5$ $h^{-1}$ kpc (even at large
off-axis angles, where the PSF size increases to $\sim 4^{''}$, the
corresponding spatial scale is only $\sim 20$ $h^{-1}$ kpc).
In any case, and ignoring for the moment the additional effect of 
the variable PSF size, the above imply that only the $w(\theta)$
of the lowest flux-limited samples could in principle be affected, 
but in the direction of reducing (and not inducing) the observed 
$\theta_0-f_{x, {\rm limit}}$ trend (since the
uncorrected $\theta_0$ values are, if anything, 
artificially larger than the true underlying one). 

However, the variability of the PSF size through-out the {\em Chandra} field
can have an additional effect, and possibly enhance or even produce
the observed $\theta_0-f_{x, {\rm limit}}$ trend.
%since in principle lower-fluxes can be observed 
%more efficiently near the center of the field while brighter ones can
%be observed equally through-out the field. 
To test for this we have repeated our analysis,
restricting the data to a circular
area of radius 6$^{'}$ around the center of the Chandra
fields, where we expect to have a relatively small variation of the PSF size. 
This choice of radius was dictated as a compromise between excluding as much
external area as possible but keeping enough sources ($\mincir$50\% of
original) to perform the clustering analysis. The results show that indeed the 
$\theta_0-f_{x, {\rm limit}}$ trend is present and qualitatively the same
as when using all the sources, implying that the
previously mentioned biases do not create the observed trend.

\subsection{Comparison with other $w(\theta)$  results}
We investigate here whether the large span of published X-ray AGN
clustering results can be explained by the derived $\theta_0-f_{x,{\rm
limit}}$ trend. To this end we attempt to take into account
the different survey area-curves, by estimating
a characteristic flux for each survey, $f_{x}(\frac{1}{2} AC)$, 
corresponding to half its area-curve 
(easy to estimate from the different survey published 
area-curves). 

In Figure 3 we plot the corresponding values of 
$\theta_0$ (for fixed $\gamma=1.8$) as a function of $f_{x}(\frac{1}{2} AC)$ 
(for both hard and soft bands) for 
the {\em Chandra} Large AREA Synoptic X-ray survey (CLASXS)
(Yang et al. 2003), the XMM/2dF (Basilakos et al. 2004; 2005),
XMM-COSMOS (Miyaji et al. 2007), XMM-ELAIS-S1 (Puccetti et al. 2006),
XMM-LSS (Gandhi et al. 2006) and AXIS (Carrera et al. 2007) surveys.
With the exception of the Yang et al. (2003) and the 
Carrera et al. (2007) hard-band results, the rest are 
consistent with the general flux-dependent trend. Note also that 
Gandhi et al. (2006) do not find any significant clustering of their 
hard-band sources. Of course, 
cosmic variance is also at work (as evidenced also by the clustering
differences between the CDF-N and CDF-S; see Fig. 2 and Gilli et al. 2005)
which should be responsible for the observed scatter around the main trend
(see also Stewart et al. 2007). 

We would like to stress that the CDF surveys have a large flux
dynamical range which is necessary in order to investigate the 
$f_{x,{\rm limit}}-\theta_o$ correlation. 
This is probably why this effect has not been clearly detected in other surveys, 
although recently, a weak such effect was found also in the CLASXS survey 
(Yang et al. 2006). 

\subsection{The spatial correlation length using $w(\theta)$}
We can use Limber's equation to invert the angular clustering and
derive the corresponding spatial clustering length, $r_0$ (eg. Peebles 1993).
To do so it is necessary to model the spatial 
correlation function as a power law and to assume a clustering evolution 
model, which we take to be that of constant clustering in comoving 
coordinates (eg. de Zotti et al. 1990; Kundi\'c 1997).
%($\epsilon=-1.2$).
%\begin{equation}
%\xi(r,z)=(r/r_{\circ})^{-\gamma} (1+z)^{-(3+\epsilon)+\gamma} \;,
%\end{equation} 
%where the parameter $\epsilon$ defines the clustering evolution model.
%Then the amplitude, $\theta_0$, of the angular correlation function
%is related to the spatial
%correlation length $r_0$ (eg. Efstathiou et al. 1991) through the equation: 
%\begin{equation}
%\theta_{\circ}^{\gamma-1}=H_{\gamma}r_{\circ}^{\gamma}
%\left(\frac{H_\circ}{c}\right)^\gamma 
%\int_{0}^{\infty} \left( \frac{1}{N}\frac{{\rm d}N}{{\rm d}z}
%\right)^{2} \frac{E(z)}{x^{\gamma-1}(z)}  %(1+z)^{-3-\epsilon+\gamma$}
%{\rm d}z \;,
%\end{equation}
%where we have used the constant clustering in comoving coordinates model
%($\epsilon=\gamma-3$). Note that $x(z)$ is the proper distance, 
%$E(z)=\sqrt{\Omega_{\rm m}(1+z)^{3}+\Omega_{\Lambda}}$ and
%$H_{\gamma}=\Gamma(\frac{1}{2}) \Gamma(\frac{\gamma-1}{2})/\Gamma
%(\frac{\gamma}{2})$. 
For the inversion to be possible it is necessary to know 
the X-ray source redshift distribution,
%, ${\rm d}N/{\rm d}z$, and the total number, $N$, of the X-ray 
%sources. Both 
which can be determined by integrating the
corresponding X-ray source luminosity function above the minimum
luminosity that corresponds to the particular flux-limit used. 
To this end we use the
Hasinger et al. (2005) and La Franca et al. (2005) LDDE luminosity functions 
for the soft and hard bands, respectively.

We perform the above inversion in the framework of the {\em
  concordance} $\Lambda$CDM cosmological model 
($\Omega_{\rm m}=1-\Omega_{\Lambda}=0.3$) %,$H_{\circ}=70$km s$^{-1}$Mpc$^{-1}$
and the comoving clustering paradigm. The resulting values of the 
spatial clustering lengths, $r_0$, show the same dependence on 
flux-limits, as in Fig.2.
 
We can compare our results with direct determinations of the spatial-correlation function 
from Gilli et al. (2005), who used 
a smaller ($\sim 50\%$) spectroscopic sample from the CDF-N and CDF-S. 
They found a significant difference between the CDF-S and CDF-N
clustering, with $r_0=10.3 \pm 1.7 \; h^{-1}$ Mpc and 
$r_0=5.5 \pm 0.6 \; h^{-1}$ Mpc, respectively (note also that the corresponding
slopes were quite shallow, roughly $\gamma\simeq 1.4-1.5$). Since, Gilli et al.
used sources from the full (0.5-8 keV) band, 
we compare their results with our soft-band results which, 
dominate the total-band sources. This comparison is possible, because 
as we have verified using a Kolmogorov-Smirnov test, the flux distributions of
the sub-samples that have spectroscopic data are statistically equivalent with
those of the whole samples.
Our inverted clustering lengths, for the lowest flux-limit used, are:
$r_0=10.3 \pm 2 \; h^{-1}$ Mpc and $r_0=6.4 \pm 2.5 \; h^{-1}$ Mpc 
(fixing $\gamma=1.8$) for the CDF-S and CDF-N respectively,
in good agreement with the Gilli et al. (2005) direct 3D determination\footnote{Leaving both
$r_{0}$ and $\gamma$ as free parameters in the fit,
%, to be fitted by the data, 
we obtain $\gamma$'s quite near their nominal value ($\gamma\sim 1.6 - 1.8$)}.

Our values can be also compared with the re-calculation of the CDF-N 
spatial clustering by Yang et al. (2006), who find 
$r_0\simeq 4.1 \pm 1.1 \; h^{-1}$ Mpc.

We return now to the strong trend between $\theta_0$ (or the 
corresponding $r_0$) and the sample flux-limit (see Fig.2 and 3),
which could be due to two possible effects 
(or the combination of both). Either the different flux-limits correspond 
to different intrinsic luminosities, ie., a luminosity-clustering dependence
(see also hints in the CLASXS and CDF-N based Yang et al. 2006 results; while for
optical data see Porciani \& Norberg 2006) or a redshift-dependent effect (ie., different 
flux-limits correspond to different redshifts traced). 
Using the sources 
which have spectroscopic redshift determinations we have derived their 
intrinsic luminosities, in each respective band, 
from their count rates using a spectral index $\Gamma=1.9$ and the
{\em concordance} $\Lambda$CDM cosmology. 
We have also applied an absorption correction by assuming a power-law 
X-ray spectrum with an intrinsic $\Gamma=1.9$, obscured by an optimum 
column density to reproduce the observed hardness ratio.
We then derive, for each flux-limit used, the median 
redshift and median luminosity of the corresponding sub-sample. We find 
relatively small variations and no monotonic change of the median redshift 
with subsample flux-limit. For example, the median spectroscopic redshift for the 
soft and hard bands, at the lowest flux-limit used, is ${\bar z}\sim 0.8$ and $\sim 0.95$,
respectively, while its mean variation between the different flux-limits used
is $\langle \delta z/z \rangle\simeq -0.11$ and -0.03 for the CDF-N and
$\langle \delta z/z \rangle \simeq -0.27$ and 0.05 for the CDF-S soft and 
hard-bands, respectively. The large redshift variation of the soft-band CDF-S data 
should be attributed to the presence of a few superclusters at $z\sim 0.7$; 
see Gilli et al. 2003).

In Figure 4 we present the correlation between the subsample
median X-ray luminosity and the corresponding subsample clustering length, as provided
by Limber's inversion. Although the CDF luminosity dynamical range is 
limited, it is evident that the median X-ray luminosity systematically increases with 
increasing sample flux-limit and it is correlated to $r_0$ (as expected from Fig.2).
It should be noted that the correlation length of the highest-flux limited CDF-S 
soft-band subsample
is by far the largest ever found ($\sim 30 h^{-1}$ Mpc), but one has to keep in mind
that the CDF-S appears not to be a typical field, as discussed earlier (see Gilli et 
al. 2003). The CDF-N high-flux results appear to converge to a value of 
$r_0 \sim 18h^{-1}$ Mpc, similar to that of some other surveys
(eg. Basilakos et al. 2004; 2005 and Puccetti et al. 2007).
%Furthermore, note that the Yang et al (2003) hard-band angular 
%clustering result  ($\theta_0 \sim 40$ arcsec) corresponds to an even larger 
%spatial clustering length. 
%Also note that the soft-band clustering amplitude of Miyaji et
%al.(2007) is $\sim 11 \;h^{-1}$ Mpc (a factor of two higher than the clustering
%amplitude of optical QSOs).Interestingly, the luminosities of the X-COSMOS sources
%are in the range $\log L = 42.5-44$. At these luminosities the high X-COSMOS 
%clustering is not very different from what we have found (see CDFN 
%results in our Figure 4).

We therefore conclude that not only are there indications for a luminosity dependent 
clustering of X-ray selected high-$z$ AGNs, but also that they are significantly more 
clustered than their lower-$z$ counterparts, which have $r_0 \sim 7-8 \; 
h^{-1}$ Mpc (eg. Akylas et al. 2000; Mullis et al. 2004). This is a clear indication of 
a strong bias evolution (eg. Basilakos, Plionis \& Ragone-Figueroa 2007).

\section{Conclusions}
We have analysed the angular clustering of the CDF-N and CDF-S X-ray AGNs and find:

\noindent
(1) A dependence of the angular clustering strength 
on the sample flux-limit. Most XMM and 
{\em Chandra} clustering analyses provide 
results that are consistent with the observed 
trend; a fact which appears to lift the confusion that arose from the 
apparent differences in their respective clustering lengths. 

\noindent
(2) Within the concordance cosmological model, the comoving clustering
evolution model and the LDDE luminosity function, our angular clustering 
results are in good agreement with direct estimations of the 
CDF-N and CDF-S spatial clustering, which are based however on 
roughly half the 
total number of sources, for which spectroscopic data were available.

\noindent 
(3) The apparent correlation between clustering strength and sample flux-limit 
transforms into a correlation between clustering strength and intrinsic
X-ray luminosity, since no significant redshift-dependent trend was found.

%\acknowledgments

\begin{table}[h]
\caption[]{The integrated angular clustering signal.}
\vspace{1cm}

\tabcolsep 15pt
\begin{tabular}{lcccc} 
Sample     & \multicolumn{2}{c}{$f_{x,{\rm limit}}$$^a$} & 
             \multicolumn{2}{c}{$f_{x,{\rm limit}}$$^b$} \\ \hline
           & $<400^{''}$   & $<900^{''}$   & $<400^{''}$  & $<900^{''}$ \\ \hline
CDF-N soft & $2.1\sigma$   & $0.0\sigma$   & $2.1\sigma$  & $1.5\sigma$ \\
CDF-N hard & $3.3\sigma$   & $2.6\sigma$   & $6.7\sigma$  & $5.5\sigma$ \\
CDF-S soft & $4.2\sigma$   & $2.6\sigma$   & $3.4\sigma$  & $3.7\sigma$ \\
CDF-S hard & $0.3\sigma$   & $1.3\sigma$   & $4.1\sigma$  & $2.7\sigma$ \\ \hline

\multicolumn{5}{l}{$^a$ the lowest flux-limit.} \\  
\multicolumn{5}{l}{$^b$ $5 \times 10^{-16}$ erg s$^{-1} $cm$^{-2}$ 
(soft band) and $10^{-15}$ erg s$^{-1} $cm$^{-2}$ (hard band).} \\

\end{tabular}
\end{table}

\clearpage

\begin{figure}
\epsscale{0.6}
\plotone{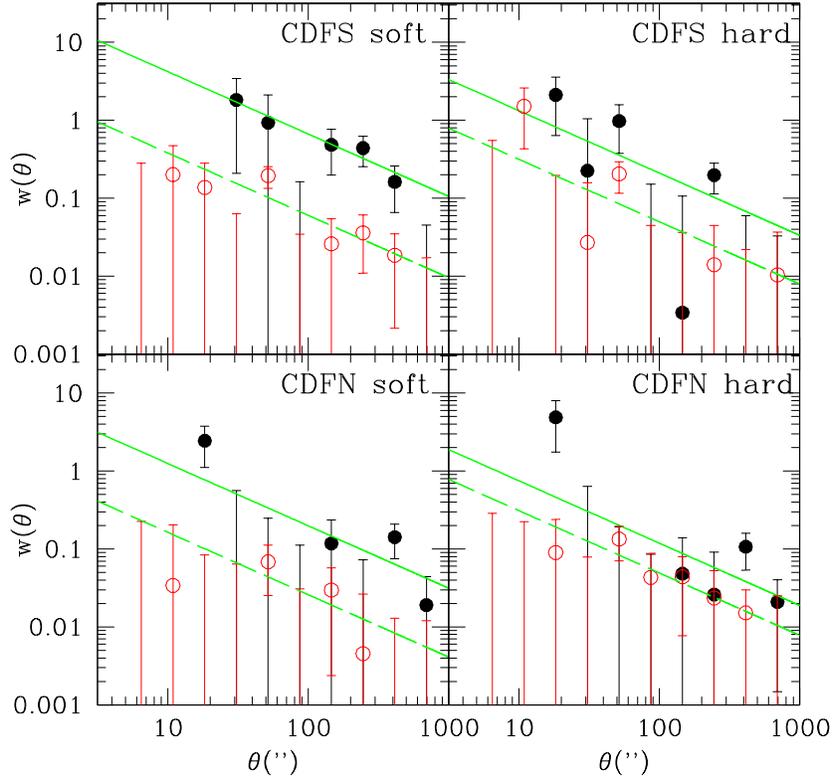}
%\plotone{wtheta.ps}
\figcaption{The CDF-S and CDF-N angular correlation function. The open points 
correspond to the overall sample, while the filled points to the highest 
flux-limit used ($f_x= 3\times 10^{-15}$ ${\rm erg/s/cm^2}$ for the soft and  
$f_x= 5\times 10^{-15}$ ${\rm erg/s/cm^2}$ 
for the hard bands, respectively). The straight 
lines correspond to the best power-law fit to the clustering
data. Errorbars correspond to $1\sigma$ Poisson uncertainties.}
\end{figure}

\begin{figure}
\epsscale{0.6}
\plotone{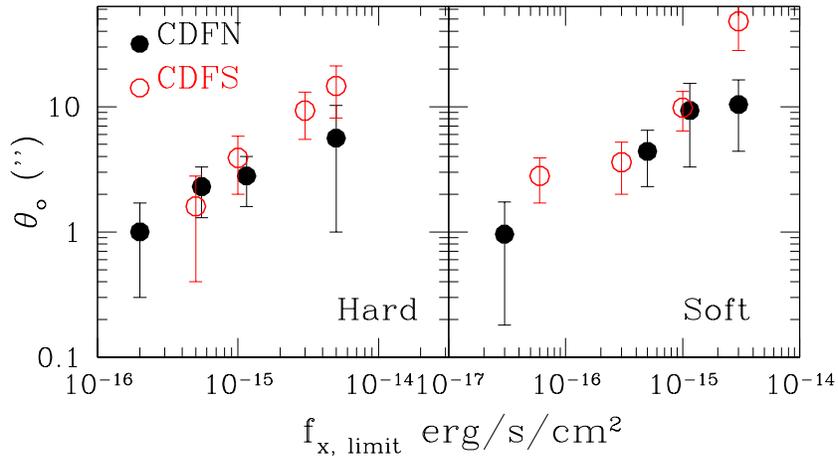}
%\plotone{theta_o1.ps}
\figcaption{The angular clustering scale as a function of the flux-limit of 
the different samples. The dependence of clustering strength to the 
flux-limit is evident. The {\em left} and {\em right} panels correspond to 
the hard and soft bands respectively. Filled symbols correspond to the CDF-N
while open ones to the CDF-S.}
\end{figure}

\begin{figure}
\epsscale{0.7}
\plotone{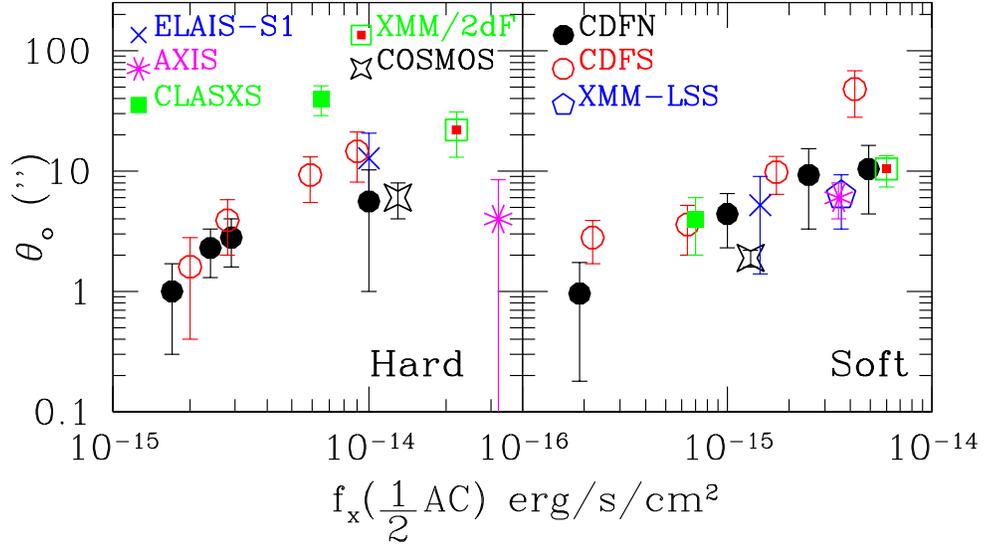}
%\plotone{theta_eff1.ps}
\figcaption{The angular correlation scale, $\theta_0$, as a function of 
different survey characteristic flux, defined as that corresponding to half the
respective survey area-curves. Most results appear to be consistent with 
the clustering flux-limit dependence, found from the CDF-N and
CDF-S. Note that in the left panel we plot the 4.5-10 keV results of
Miyaji et al. (2007). Errorbars correspond to $1\sigma$ Poisson uncertainties.}
\end{figure}

\begin{figure}
\epsscale{0.7}
\plotone{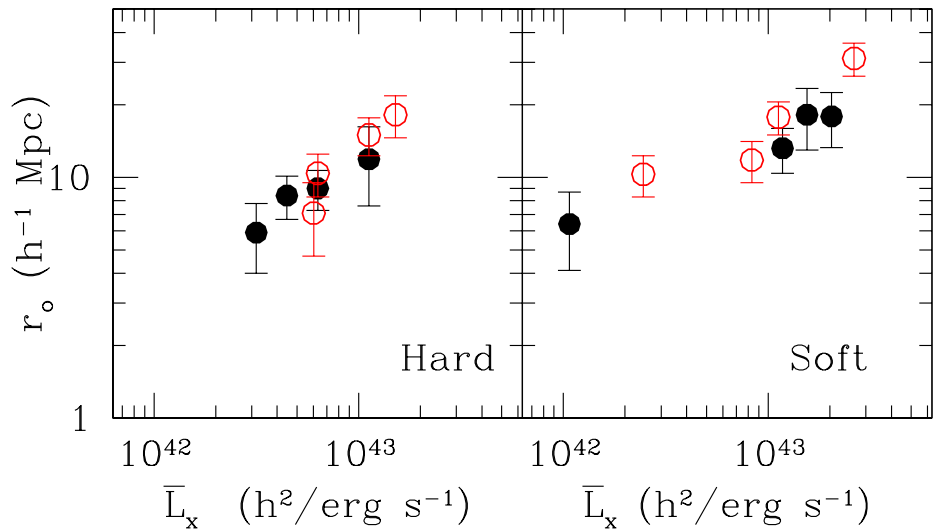}
\figcaption{The correlation between the clustering length, $r_0$, and the 
median intrinsic X-ray luminosity of each subsample.
The {\em left} and {\em right} panels correspond to 
the hard and soft bands respectively. Filled symbols correspond to the CDF-N
while open ones to the CDF-S.}
\end{figure}

\end{document}